\documentclass[runningheads]{svmult}
\usepackage{makeidx}
\usepackage{graphicx}
\usepackage{subeqnar}
\usepackage{multicol}
\usepackage{physprbb}
\makeindex 

\def\AGUT{{}\;\;\raisebox{.9ex}{$\times$}\raisebox{-.5ex}%
{$\!\!\!\!\!\!\!\!\scriptscriptstyle i=1,2,3$} \,(SMG_i \times %
U(1)_{\scriptscriptstyle B-L,i})}
\def\sVEV#1{\left\langle #1\right\rangle}
\def\abs#1{\left| #1\right|}
\def\sleq{\raisebox{-.6ex}{${\textstyle\stackrel{<}{\sim}}$}}

\begin{document}
\title*{Family replicated fit of all quark and lepton masses and mixings\footnote{NBI-HE-02-01}}
\toctitle{8th Adriatic Meeting -- Particle Physics in the New Millennium}
\titlerunning{Family replicated fit of all quark and lepton masses and mixings}
\author{H.~B.~Nielsen \and Y.~Takanishi}
\authorrunning{H.~B.~Nielsen and Y.~Takanishi}
\institute{Deutsches Elektronen-Synchrotron DESY, \\
Notkestra{\ss}e 85, D-22603 Hamburg, Germany \\
\mbox{~}\\
The Niels Bohr Institute, \\
Blegdamsvej 17, DK-2100 Copenhagen {\O}, Denmark}

\maketitle

\begin{abstract}

We review our recent development of family replicated 
gauge group model, which generates the Large Mixing Angle MSW 
solution. The model is based on each family of quarks and leptons 
having its own set of gauge fields, each containing a replica of 
the Standard Model gauge fields plus a $(B-L)$-coupled gauge 
field. A fit of all the seventeen quark-lepton mass and mixing 
angle observables, using just six new Higgs field vacuum 
expectation values, agrees with the experimental data 
order of magnitudewise. However, this model can not
predict the baryogenesis in right order, therefore, we
discuss further modification of our model and present
a preliminary result of baryon number to entropy ratio.

\end{abstract}

\section{Introduction}

We have previously attempted to fit all the fermion masses and their 
mixing angles~\cite{FNT,NT1} including 
baryogenesis~\cite{NT2} in a model without
supersymmetry or grand unification. 
This model has the 
maximum number of gauge fields consistent with maintaining 
the irreduciblity of the usual Standard Model fermion 
representations, added three right-handed neutrinos. 
The predictions of this previous model are in order 
of magnitude agreement with all existing experimental data, 
however, only provided we use the Small Mixing Angle MSW~\cite{MSW} (SMA-MSW) 
solution. But, for 
the reasons given below, the SMA-MSW solution is now 
disfavoured by experiments. So here we review a modified 
version of the previous model, which 
manages to accommodate the Large Mixing Angle MSW (LMA-MSW) solution 
for solar neutrino oscillations using 6 additional Higgs fields (relative 
to the Standard Model) vacuum 
expectation values (VEVs) as adjustable parameters.

A neutrino oscillation solution to the solar neutrino problem
and a favouring of the LMA-MSW solution 
is supported by SNO results~\cite{SNO}: The measurement of the $^8$B and $hep$ 
solar neutrino fluxes shows no significant energy dependence 
of the electron neutrino survival probability in the
Super-Kamiokande and SNO energy ranges. 

Moreover, the important result which also supports LMA-MSW solution 
on the solar neutrino problem, 
reported by the Super-Kamiokande collaboration~\cite{SKDN}, 
that the day-night asymmetry data disfavour the 
SMA-MSW solution at the $95\%$ C.L..

In fact, global analyses~\cite{fogli,cc1,goswami,smirnov} of all
solar neutrino data have confirmed that the LMA-MSW solution gives the best 
fit to the data and that the SMA-MSW solution is very strongly
disfavoured and only acceptable at the $3\sigma$ level. Typical best fit
values of the mass squared difference and mixing angle parameters 
in the two flavour LMA-MSW solution are 
$\Delta m^2_\odot\approx4.5\times 10^{-5}~\mbox{\rm eV}^2$ 
and $\tan^2\theta_{\odot}\approx0.35$.

This paper is organised as follows: In the next section, we 
present our gauge group -- the family replicated gauge group -- 
and the quantum numbers of fermion and Higgs fields. 
Then, in section $3$ we discuss our philosophy of all 
gauge- and Yukawa couplings
at Planck scale being of order unity. In section 
$4$ we address how 
the family replicated gauge group breaks down to Standard Model gauge
group, and we add a small review of see-saw mechanism.
The mass matrices of all sectors are presented in section $5$,
the renormalisation group equations -- renormalisable and also 
5 dimensional non-renormalisable ones -- are shown in section $6$.
The calculation 
is described in section $7$ and the results are presented 
in section $8$. We discuss further modification of our model 
and present a preliminary results of baryon number to entropy ratio
in section $9$. Finally, section $10$ contains our conclusion.

\section{Quantum numbers of model}

Our model has, as its back-bone, the property that there are generations 
(or families) not only for fermions but also 
for the gauge bosons, \mbox{{\it i.e.}}, we have a generation (family) 
replicated gauge group namely 
\begin{equation}
  \label{eq:frg}
  \AGUT \hspace{2mm},
\end{equation}
where $SMG$ denotes the Standard Model gauge group 
$\equiv SU(3)\times SU(2)\times U(1)$, $\times$ denotes the 
Cartesian product and $i$ runs through the generations, $i=1,2,3$.

Note that this family replicated gauge group, eq.~(\ref{eq:frg}), 
is the maximal gauge group under the following assumptions:
\begin{itemize}
\item It should only contain transformations which change the known 
45 (= 3 generations of 15 Weyl particles each) Weyl fermions of the Standard Model 
and the additional three heavy see-saw (right-handed) neutrinos. 
That is our gauge group is assumed to be a subgroup of $U(48)$.
\item We avoid any new gauge transformation that would transform a 
Weyl state from one irreducible representation of the Standard Model 
group into another irreducible representation: 
there is no gauge coupling unification.
\item  The gauge group does not contain any anomalies in the gauge 
symmetry -- neither gauge nor mixed anomalies even without using the 
Green-Schwarz anomaly cancelation mechanism.
\item It should be as big as possible under the foregoing assumptions.
\end{itemize}

\begin{table}[!t]
\caption{All $U(1)$ quantum charges in the family replicated model. 
The symbols for the fermions shall be considered to mean
``proto''-particles. Non-abelian representations are given by a rule 
from the abelian ones (see Eq.~(\ref{eq:mod})).}
\vspace{3mm}
\label{Table1}
\begin{center}
\begin{tabular}{|c||c|c|c|c|c|c|} \hline
& $SMG_1$& $SMG_2$ & $SMG_3$ & $U_{\scriptscriptstyle B-L,1}$ & 
$U_{\scriptscriptstyle B-L,2}$ & $U_{\scriptscriptstyle B-L,3}$ \\ \hline\hline
$u_L,d_L$ &  $\frac{1}{6}$ & $0$ & $0$ & $\frac{1}{3}$ & $0$ & $0$ \\
$u_R$ &  $\frac{2}{3}$ & $0$ & $0$ & $\frac{1}{3}$ & $0$ & $0$ \\
$d_R$ & $-\frac{1}{3}$ & $0$ & $0$ & $\frac{1}{3}$ & $0$ & $0$ \\
$e_L, \nu_{e_{\scriptscriptstyle L}}$ & $-\frac{1}{2}$ & $0$ & $0$ & $-1$ & $0$ 
& $0$ \\
$e_R$ & $-1$ & $0$ & $0$ & $-1$ & $0$ & $0$ \\
$\nu_{e_{\scriptscriptstyle R}}$ &  $0$ & $0$ & $0$ & $-1$ & $0$ & $0$ \\ \hline
$c_L,s_L$ & $0$ & $\frac{1}{6}$ & $0$ & $0$ & $\frac{1}{3}$ & $0$ \\
$c_R$ &  $0$ & $\frac{2}{3}$ & $0$ & $0$ & $\frac{1}{3}$ & $0$ \\
$s_R$ & $0$ & $-\frac{1}{3}$ & $0$ & $0$ & $\frac{1}{3}$ & $0$\\
$\mu_L, \nu_{\mu_{\scriptscriptstyle L}}$ & $0$ & $-\frac{1}{2}$ & $0$ & $0$ & 
$-1$ & $0$\\
$\mu_R$ & $0$ & $-1$ & $0$ & $0$  & $-1$ & $0$ \\
$\nu_{\mu_{\scriptscriptstyle R}}$ &  $0$ & $0$ & $0$ & $0$ & $-1$ & $0$ \\ \hline
$t_L,b_L$ & $0$ & $0$ & $\frac{1}{6}$ & $0$ & $0$ & $\frac{1}{3}$ \\
$t_R$ &  $0$ & $0$ & $\frac{2}{3}$ & $0$ & $0$ & $\frac{1}{3}$ \\
$b_R$ & $0$ & $0$ & $-\frac{1}{3}$ & $0$ & $0$ & $\frac{1}{3}$\\
$\tau_L, \nu_{\tau_{\scriptscriptstyle L}}$ & $0$ & $0$ & $-\frac{1}{2}$ & $0$ & 
$0$ & $-1$\\
$\tau_R$ & $0$ & $0$ & $-1$ & $0$ & $0$ & $-1$\\
$\nu_{\tau_{\scriptscriptstyle R}}$ &  $0$ & $0$ & $0$ & $0$ & $0$ & $-1$ \\ 
\hline \hline
$\phi_{\scriptscriptstyle WS}$ & $0$ & $\frac{2}{3}$ & $-\frac{1}{6}$ & $0$ & 
$\frac{1}{3}$ & $-\frac{1}{3}$ \\
$\omega$ & $\frac{1}{6}$ & $-\frac{1}{6}$ & $0$ & $0$ & $0$ & $0$\\
$\rho$ & $0$ & $0$ & $0$ & $-\frac{1}{3}$ & $\frac{1}{3}$ & $0$\\
$W$ & $0$ & $-\frac{1}{2}$ & $\frac{1}{2}$ & $0$ & $-\frac{1}{3}$ & $\frac{1}{3}$ \\
$T$ & $0$ & $-\frac{1}{6}$ & $\frac{1}{6}$ & $0$ & $0$ & $0$\\
$\chi$ & $0$ & $0$ & $0$ & $0$ & $-1$ & $1$ \\
$\phi_{\scriptscriptstyle B-L}$ & $0$ & $0$ & $0$ & $0$ & $0$ & $2$ \\ 
\hline
\end{tabular}
\end{center}
\end{table}

The quantum numbers of the particles/fields in our model are found in table 1
and use of the following procedure: In table 1 one finds the charges 
under the six $U(1)$ groups in the gauge 
group \ref{eq:frg}.
Then for each particle one should take the representation under the $SU(2)_i$
and $SU(3)_i$ groups ($i=1,2,3$) with
lowest dimension matching to $y_i/2$ according to the requirement 
\begin{equation}
  \label{eq:mod}
  \frac{t_i}{3}+\frac{d_i}{2} + \frac{y_i}{2} = 0~~{\rm (mod~1)}\hspace{2mm},
\end{equation}
where $t_i$ and $d_i$ are the triality and duality for 
the $i$'th proto-generation gauge groups $SU(3)_i$ and $SU(2)_i$ 
respectively.

\section{The philosophy of all couplings being order unity}
\label{philo}

Any realistic model and at least certainly our model tends to 
get far more fundamental couplings than we have parameters 
in the Standard Model and thus pieces of data to fit. This 
is especially so for our model based on many $U(1)$ charges~\cite{FN} 
because we take it to have 
practically any not mass protected particles one may propose 
at the fundamental mass scale, taken to be the Planck mass. Especially 
we assume the existence of Dirac fermions with order of fundamental 
scale masses needed to allow the quark and lepton Weyl particles to 
take up successively gauge charges from the Higgs fields VEVs.
So unless we make assumptions about the many coupling constants 
and fundamental masses we have no chance to predict anything. Almost 
the only chance of making an assumption about all these couplings, 
which is not very model dependent, is to assume that they are {\em all of 
order unity} in the fundamental unit. This is the same type of assumption 
that is really behind use of dimensional arguments to estimate sizes of 
quantities. A procedure very often used successfully. If we really 
assumed every coupling and mass of order unity we would get the 
effective Yukawa couplings of the quarks and leptons to the 
Weinberg-Salam Higgs field to be also of order unity what is 
phenomenologically not true. To avoid this prediction we then 
blame the smallness of all but the top-Yukawa coupling on smallness 
in fundamental Higgs VEVs. That is to say we assume that the 
VEVs of the Higgs fields in Table 1, $\rho$, $\omega$, $T$, 
$W$, $\chi$,
$\phi_{B-L}$ and $\phi_{WS}$ are (possibly) very small 
compared to the fundamental/Planck unit, and these are the 
quantities we have to fit.

Technically we implement these unknown -- but of order unity according to 
our assumption -- couplings and masses by a Monte Carlo technique: we put 
them equal to random numbers with a distribution dominated by numbers 
of order unity and then perform the calculation of the observable 
quantities such as quark or lepton masses and mixing
angles again and again. At the end we average the logarithmic of 
these quantities and exponentiate them. In this way we expect to get the
typical order of magnitude predicted for the observable quantities.
In praxis we do not have to put random numbers in for all the many 
couplings in the fundamental model, but can instead just provide each 
mass matrix element with a single random number factors.

After all a product of several of order unity factors is just an order 
unity factor again. To resume our model philosophy:
{\em Only Higgs field VEVs are not of order unity. We must be
 satisfied with order of magnitude results.}

\section{Breaking of the family replicated gauge group to the Standard Model}
\indent\ The family replicated gauge group broken down to its diagonal
subgroup at scales about one or half order of magnitude under the Planck 
scale by Higgs fields -- $W$, $T$, $\omega$, $\rho$ 
and $\chi$ (in Table~\ref{Table1}):
\begin{equation}
  \label{eq:subagut}
\AGUT \rightarrow SMG\times U(1)_{B-L}\,.
\end{equation}
This diagonal subgroup is further broken down by yet two more 
Higgs fields --- the Weinberg-Salam Higgs field $\phi_{WS}$ and 
another Higgs field $\phi_{B-L}$ --- to 
$SU(3)\times U(1)_{em}$. 

\subsection{See-saw mechanism}
\indent\ See-saw mechanics is build into our model to fit the scale 
of the neutrino oscillations, \mbox{\it i.e.},  we use the 
right-handed neutrinos with heavy Majorana masses ($10^{11}$ GeV).

In order to mass-protect the right-handed neutrino from getting Planck scale
masses, we have to introduce $\phi_{\scriptscriptstyle B-L}$ 
which breaks the $B-L$ quantum charge spontaneously, and using this
new Higgs filed we are able to deal the neutrino oscillations,
\mbox{\it i.e.}, to fit the scale of the see-saw particle masses.
However, due to mass-protection by the Standard Model gauge symmetry, the 
left-handed Majorana mass terms should be negligible in our model.
Then, naturally, the light neutrino mass matrix -- effective left-left 
transition Majorana mass matrix -- can be obtained via the see-saw 
mechanism~\cite{seesaw}:
\begin{equation}
  \label{eq:meff}
  M_{\rm eff} \! \approx \! M^D_\nu\,M_R^{-1}\,(M^D_\nu)^T\,.
\end{equation}

\section{Mass matrices}

Using the $U(1)$ fermion quantum charges and Higgs field 
(presented in Table~\ref{Table1}) we can calculate the degrees of suppressions
of the left-right transition -- Dirac mass -- matrices and 
also Majorana mass matrix (right-right transition).

Note that the random complex order of unity 
numbers which are supposed to multiply all the mass matrix elements 
are not represented in following matrices:
\noindent
the up-type quarks:
\begin{eqnarray}
M_{\scriptscriptstyle U} \simeq \frac{\sVEV{(\phi_{\scriptscriptstyle\rm WS})^\dagger}}{\sqrt{2}}
\hspace{-0.1cm}
\left(\!\begin{array}{ccc}
        (\omega^\dagger)^3 W^\dagger T^2
        & \omega \rho^\dagger W^\dagger T^2
        & \omega \rho^\dagger (W^\dagger)^2 T\\
        (\omega^\dagger)^4 \rho W^\dagger T^2
        &  W^\dagger T^2
        & (W^\dagger)^2 T\\
        (\omega^\dagger)^4 \rho
        & 1
        & W^\dagger T^\dagger
\end{array} \!\right)\label{M_U}
\end{eqnarray}  
\noindent
the down-type quarks:
\begin{eqnarray}
M_{\scriptscriptstyle D} \simeq \frac{\sVEV{\phi_{\scriptscriptstyle\rm WS}}}
{\sqrt{2}}\hspace{-0.1cm}
\left (\!\begin{array}{ccc}
        \omega^3 W (T^\dagger)^2
      & \omega \rho^\dagger W (T^\dagger)^2
      & \omega \rho^\dagger T^3 \\
        \omega^2 \rho W (T^\dagger)^2
      & W (T^\dagger)^2
      & T^3 \\
        \omega^2 \rho W^2 (T^\dagger)^4
      & W^2 (T^\dagger)^4
      & W T
                        \end{array} \!\right) \label{M_D}
\end{eqnarray}
\noindent %
the charged leptons:
\begin{eqnarray}        
M_{\scriptscriptstyle E} \simeq \frac{\sVEV{\phi_{\scriptscriptstyle\rm WS}}}
{\sqrt{2}}\hspace{-0.1cm}
\left(\hspace{-0.1 cm}\begin{array}{ccc}
    \omega^3 W (T^\dagger)^2
  & (\omega^\dagger)^3 \rho^3 W (T^\dagger)^2 
  & (\omega^\dagger)^3 \rho^3 W T^4 \chi \\
    \omega^6 (\rho^\dagger)^3  W (T^\dagger)^2 
  &   W (T^\dagger)^2 
  &  W T^4 \chi\\
    \omega^6 (\rho^\dagger)^3  (W^\dagger)^2 T^4 
  & (W^\dagger)^2 T^4
  & WT
\end{array} \hspace{-0.1cm}\right) \label{M_E}
\end{eqnarray}
\noindent
the Dirac neutrinos:
\begin{eqnarray}
M^D_\nu \simeq \frac{\sVEV{(\phi_{\scriptscriptstyle\rm WS})^\dagger}}{\sqrt{2}}
\hspace{-0.1cm}
\left(\hspace{-0.1cm}\begin{array}{ccc}
        (\omega^\dagger)^3 W^\dagger T^2
        & (\omega^\dagger)^3 \rho^3 W^\dagger T^2
        & (\omega^\dagger)^3 \rho^3 W^\dagger  T^2 \chi\\
        (\rho^\dagger)^3 W^\dagger T^2
        &  W^\dagger T^2
        & W^\dagger T^2 \chi\\
        (\rho^\dagger)^3 W^\dagger T^\dagger \chi^\dagger
        &  W^\dagger T^\dagger \chi^\dagger
        & W^\dagger T^\dagger
\end{array} \hspace{-0.1 cm}\right)\label{Mdirac}
\end{eqnarray} 
\noindent %
and the Majorana (right-handed) neutrinos:
\begin{eqnarray}    
M_R \simeq \sVEV{\phi_{\scriptscriptstyle\rm B-L}}\hspace{-0.1cm}
\left (\hspace{-0.1 cm}\begin{array}{ccc}
(\rho^\dagger)^6 (\chi^\dagger)^2
& (\rho^\dagger)^3 (\chi^\dagger)^2
& (\rho^\dagger)^3 \chi^\dagger \\
(\rho^\dagger)^3 (\chi^\dagger)^2
& (\chi^\dagger)^2 & \chi^\dagger \\
(\rho^\dagger)^3 \chi^\dagger & \chi^\dagger
& 1
\end{array} \hspace{-0.1 cm}\right ) \label{Mmajo}
\end{eqnarray}

\section{Renormalisation group equations from Planck scale to week scale 
via see-saw scale}
\indent\ It should be kept in mind that the effective Yukawa couplings for 
the Weinberg-Salam Higgs field, which 
are given by the Higgs field factors in the above mass matrices 
multiplied by order unity factors, 
are the running Yukawa couplings at a scale {\em near the Planck 
scale}. In this way, we had to 
use the renormalisation group (one-loop)
$\beta$-functions to run these couplings down to the experimentally 
observable scale which we took for the charged fermion masses to be 
compared to ``measurements'' 
at the scale of $1~\mbox{\rm GeV}$, except for the 
top quark mass prediction. We define the top quark pole mass:
\begin{equation}
M_t = m_t(M)\left(1+\frac{4}{3}\frac{\alpha_s(M)}{\pi}\right)\hspace{2mm},
\end{equation}
where we put $M=180~\mbox{\rm GeV}$ for simplicity.

We use the one-loop $\beta$ functions for the gauge 
couplings and the charged fermion 
Yukawa matrices~\cite{pierre} as follows:
\begin{eqnarray}
  \label{eq:recha}
16 \pi^2 {d g_{1}\over d  t} &\!=\!& \frac{41}{10} \, g_1^3 \nonumber\\
16 \pi^2 {d g_{2}\over d  t} &\!=\!& - \frac{19}{16} \, g_2^3  \nonumber\\
16 \pi^2 {d g_{3}\over d  t} &\!=\!& - 7 \, g_3^3  \nonumber\\
16 \pi^2 {d Y_{\scriptscriptstyle U}\over d  t} &\!=\!& \frac{3}{2}\, 
\left( Y_{\scriptscriptstyle U} (Y_{\scriptscriptstyle U})^\dagger
-  Y_{\scriptscriptstyle D} (Y_{\scriptscriptstyle D})^\dagger\right)\, Y_{\scriptscriptstyle U} \nonumber\\
&& + \left\{\, Y_{\scriptscriptstyle S} - \left(\frac{17}{20} g_1^2 
+ \frac{9}{4} g_2^2 + 8 g_3^2 \right) \right\}\, Y_{\scriptscriptstyle U}\\
16 \pi^2 {d Y_{\scriptscriptstyle D}\over d  t} &\!=\!& \frac{3}{2}\, 
\left( Y_{\scriptscriptstyle D} (Y_{\scriptscriptstyle D})^\dagger
-  Y_{\scriptscriptstyle U} (Y_{\scriptscriptstyle U})^\dagger\right)\,Y_{\scriptscriptstyle D} \nonumber\\ 
&& + \left\{\, Y_{\scriptscriptstyle S} - \left(\frac{1}{4} g_1^2 
+ \frac{9}{4} g_2^2 + 8 g_3^2 \right) \right\}\, Y_{\scriptscriptstyle D} \nonumber\\
16 \pi^2 {d Y_{\scriptscriptstyle E}\over d  t} &\!=\!& \frac{3}{2}\, 
\left( Y_{\scriptscriptstyle E} (Y_{\scriptscriptstyle E})^\dagger \right)\,Y_{\scriptscriptstyle E} \nonumber\\
&& + \left\{\, Y_{\scriptscriptstyle S} - \left(\frac{9}{4} g_1^2 
+ \frac{9}{4} g_2^2 \right) \right\}\, Y_{\scriptscriptstyle E}\nonumber\\
Y_{\scriptscriptstyle S} &\!=\!& {{\rm Tr}{}}(\, 3\, Y_{\scriptscriptstyle U}^\dagger\, Y_{\scriptscriptstyle U} 
+  3\, Y_{\scriptscriptstyle D}^\dagger \,Y_{\scriptscriptstyle D} +  Y_{\scriptscriptstyle E}^\dagger\, Y_{\scriptscriptstyle E}\,)  \nonumber\hspace{2mm},
\end{eqnarray}
where $t=\ln\mu$.

By calculation we use the following initial values of gauge coupling constants:
\begin{eqnarray}
U(1): \quad & g_1(M_Z) = 0.462\hspace{2mm}, \quad & g_1(M_{\rm Planck}) = 0.614\\
SU(2):\quad & g_2(M_Z) = 0.651\hspace{2mm}, \quad & g_2(M_{\rm Planck}) = 0.504\\
SU(3):\quad & g_3(M_Z) = 1.22 \hspace{2mm}, \quad & g_3(M_{\rm Planck}) = 0.491
\end{eqnarray}

\subsection{The renormalisation group equations for the effective 
neutrino mass matrix}
\indent

The effective light neutrino masses are given by an irrelevant, 
nonrenormalisable (5 dimensional term) -- 
effective mass matrix $M_{\rm eff}$ -- for which the running
formula is~\cite{Meffrun}:
\begin{equation}
\label{eq:remeff}
16 \pi^2 {d M_{\rm eff} \over d  t}
= ( - 3 g_2^2 + 2 \lambda + 2 Y_{\scriptscriptstyle S} ) \,M_{\rm eff}
- {3\over 2} \left( M_{\rm eff}\, ( Y_{\scriptscriptstyle E} Y_{\scriptscriptstyle E}^\dagger )^T 
+ ( Y_{\scriptscriptstyle E} Y_{\scriptscriptstyle E}^\dagger ) \,M_{\rm eff}\right) \hspace{2mm},
\end{equation}
where $\lambda$ is the Weinberg-Salam Higgs self-coupling constant and
the mass of the Standard Model Higgs boson is given by 
$M_H^2 = \lambda \sVEV{\phi_{WS}}^2$. We just for simplicity 
take $M_H = 115~\mbox{\rm GeV}$ thereby we ignore the running of
the Higgs self-coupling and fixed as $\lambda=0.2185$.

Note that the renormalisation group equations are used to evolve 
the effective neutrino mass matrix from the see-saw 
sale, set by $\sVEV{\phi_{B-L}}$ in our model, to $1~\mbox{\rm GeV}$. 

\section{Method of numerical computation}
\label{sec:FN}
\indent\ In the philosophy of order unity numbers spelled out in
section~\ref{philo} we evaluate 
the product of mass-protecting Higgs VEVs required 
for each mass matrix element and provide it  
with a random complex number, $\lambda_{ij}$, of order one as a factor
taken to have Gaussian distribution with mean value zero. But we
hope the exact form of distribution does not matter much provided 
we have $\sVEV{\ln \abs{\lambda_{ij}}}=0$. In this way, we 
simulate a long chain of fundamental Yukawa couplings 
and propagators making the transition corresponding to an 
effective Yukawa coupling in the Standard Model and the parameters 
in neutrino sector. In the numerical 
computation we then calculate the masses and mixing angles time 
after time, using different sets of random numbers and, in the 
end, we take the logarithmic average of the calculated quantities 
according to the following formula:
\begin{equation}
  \label{eq:avarage}
  \sVEV{m}=\exp\left(\sum_{i=1}^{N} \frac{\ln m_i}{N}\right) \,. 
\end{equation}
Here $\sVEV{m}$ is what we take to be the prediction for one of the 
masses or mixing angles, $m_i$ is the result of the calculation
done with one set of random number combinations and $N$ is the total 
number of random number combinations used.

Since we only expect to make order of magnitude fits, we should of 
course not use ordinary $\chi^2$ defined form the experimental 
uncertainties by rather the $\chi^2$ that would correspond to a 
relative uncertainly -- an uncertain factor of order unity. Since
the normalisation of such a $\chi^2$ is not so easy to choose 
exactly we define instead a quantity which we call the 
goodness of fit (\mbox{\rm g.o.f.}). Since 
our model can only make predictions order of magnitudewise, this 
quantity \mbox{\rm g.o.f.} should only depend on the ratios of
the fitted masses and mixing angles to the experimentally
determined masses and mixing angles:
\begin{equation}
\label{gof}
\mbox{\rm g.o.f.}\equiv\sum \left[\ln \left(
\frac{\sVEV{m}}{m_{\rm exp}} \right) \right]^2 \,,
\end{equation}
where $m_{\rm exp}$ are the
corresponding experimental values presented in 
Table~\ref{convbestfit}.

We should emphasise that we \underline{do not} adjust
the order of one numbers by selection, \mbox{\it i.e.}, the
complex random numbers are needed for only calculational 
purposes. That means that we have only six adjustable parameters
-- VEVs of Higgs fields -- and, on the other hand, that the 
averages of the predicted quantities, $\sVEV{m}$, are just 
results of integration
over the ``dummy'' variables -- random numbers -- therefore, the
random numbers are not at all parameters!

Strictly speaking, however, one could consider the choice of the 
distribution of the random order unity numbers as parameters. But 
we hope that provided we impose on the distribution the conditions that 
the average be zero and the average of the logarithm of the numerical
value be zero, too, any reasonably smooth distribution would 
give similar results for the $\sVEV{m}$ values at the end. In our 
early work~\cite{NT1} we did see that a couple of different proposals 
did not make too much difference.

\section{Results}
\indent\ We averaged over $N=10,000$ complex order unity random 
number combinations. These complex numbers
are chosen to be a number picked from a Gaussian 
distribution, with mean value zero and standard deviation one, 
multiplied by a random phase factor. We put them as factors into 
the mass matrices (eqs.~\ref{M_U}-\ref{Mmajo}). Then 
we computed averages according to eq.~(\ref{eq:avarage}) and used 
eq.~(\ref{gof}) as a $\chi^2$ to fit the $6$ free parameters and found:
\begin{eqnarray} 
\label{eq:VEVS} 
&&\sVEV{\phi_{\scriptscriptstyle WS}}= 246~\mbox{\rm GeV}\,,  
\,\sVEV{\phi_{\scriptscriptstyle B-L}}=1.64\times10^{11}~\mbox{\rm GeV}\,, 
\,\sVEV{\omega}=0.233\,,\nonumber\\
&&\,\sVEV{\rho}=0.246\,,\,\sVEV{W}=0.134\,,
\,\sVEV{T}=0.0758\,,\,\sVEV{\chi}=0.0737\,,
\end{eqnarray}
where, except for the Weinberg-Salam Higgs field and 
$\sVEV{\phi_{\scriptscriptstyle B-L}}$, the VEVs are expressed in Planck units. 
Hereby we have considered that the Weinberg-Salam Higgs field VEV is 
already fixed by the Fermi constant.
The results of the best fit, with the VEVs in eq.~(\ref{eq:VEVS}), 
are shown in Table~\ref{convbestfit} and the fit has  
$\mbox{\rm g.o.f.}=3.63$. 

We have $11=17 - 6$ degrees of freedom -- predictions -- leaving each of 
them with a logarithmic error of
$\sqrt{3.63/11}\simeq0.57$, which is very close to the 
theoretically expected value $64\%$~\cite{FF}. This means 
that we can fit {\rm all quantities} within a factor 
$\exp\left(\sqrt{3.63/11}\right)\simeq1.78$ of the experimental values.

\begin{table}[!t]
\caption{Best fit to conventional experimental data.
All masses are running
masses at $1~\mbox{\rm GeV}$ except the top quark mass which is the pole mass.
Note that we use the square roots of the neutrino data in this 
Table, as the fitted neutrino mass and mixing parameters 
$\sVEV{m}$, in our goodness of fit ($\mbox{\rm g.o.f.}$) definition, 
eq.~(\ref{gof}).}
\begin{displaymath}
\begin{array}{|c|c|c|}
\hline\hline
 & {\rm Fitted} & {\rm Experimental} \\ \hline
m_u & 4.4~\mbox{\rm MeV} & 4~\mbox{\rm MeV} \\
m_d & 4.3~\mbox{\rm MeV} & 9~\mbox{\rm MeV} \\
m_e & 1.0~\mbox{\rm MeV} & 0.5~\mbox{\rm MeV} \\
m_c & 0.63~\mbox{\rm GeV} & 1.4~\mbox{\rm GeV} \\
m_s & 340~\mbox{\rm MeV} & 200~\mbox{\rm MeV} \\
m_{\mu} & 80~\mbox{\rm MeV} & 105~\mbox{\rm MeV} \\
M_t & 208~\mbox{\rm GeV} & 180~\mbox{\rm GeV} \\
m_b & 7.2~\mbox{\rm GeV} & 6.3~\mbox{\rm GeV} \\
m_{\tau} & 1.1~\mbox{\rm GeV} & 1.78~\mbox{\rm GeV} \\
V_{us} & 0.093 & 0.22 \\
V_{cb} & 0.027 & 0.041 \\
V_{ub} & 0.0025 & 0.0035 \\ \hline
\Delta m^2_{\odot} & 9.5 \times 10^{-5}~\mbox{\rm eV}^2 &  4.5 \times 10^{-5}~\mbox{\rm eV}^2 \\
\Delta m^2_{\rm atm} & 2.6 \times 10^{-3}~\mbox{\rm eV}^2 &  3.0 \times 10^{-3}~\mbox{\rm eV}^2\\
\tan^2\theta_{\odot} &0.23 & 0.35\\
\tan^2\theta_{\rm atm}& 0.65 & 1.0\\
\tan^2\theta_{13}  & 4.8 \times 10^{-2} & \sleq~2.6 \times 10^{-2}\\
\hline\hline
\mbox{\rm g.o.f.} &  3.63 & - \\
\hline\hline
\end{array}
\end{displaymath}
\label{convbestfit}
\end{table}

{}From the table~\ref{convbestfit} the experimental mass values are a factor
two higher than predicted for down, charm and for the Cabibbo angle
$V_{us}$ while they are smaller by a factor for strange and 
electron. Thinking only on the angles and masses (not squared) the agreement
is in other cases better than a factor two.

Experimental results for the values of neutrino mixing angles 
are often presented in terms of the function $\sin^22\theta$ 
rather than $\tan^2\theta$ (which, contrary to $\sin^22\theta$, 
does not have a maximum at $\theta=\pi/4$ and thus still varies 
in this region).
Transforming from $\tan^2\theta$ variables to $\sin^22\theta$ 
variables, our predictions for the neutrino mixing angles become:
\begin{eqnarray}
  \label{eq:sintan}
 \sin^22\theta_{\odot} &\!=\!&0.61\,,\\
 \sin^22\theta_{\rm atm} &\!=\!& 0.96\,, \\
 \sin^22\theta_{13} &\!=\!& 0.17\,.
\end{eqnarray}  
We also give here our predicted hierarchical neutrino mass 
spectrum:
\begin{eqnarray}
m_1 &\!=\!&  4.9\times10^{-4}~~\mbox{\rm eV}\,, 
\label{eq:neutrinomass1}\\
m_2 &\!=\!&  9.7\times10^{-3}~~\mbox{\rm eV}\,, 
\label{eq:neutrinomass2}\\
m_3 &\!=\!&  5.2\times10^{-2}~~\mbox{\rm eV}\,.
\label{eq:neutrinomass3} 
\end{eqnarray}

Our agreement with experiment is excellent: all of our 
order of magnitude neutrino predictions lie 
inside the $99\%$ C.L. border determined from phenomenological fits 
to the neutrino data, even including the CHOOZ upper bound.
Our prediction of the solar mass squared 
difference is about a factor of $2$ larger than the global data fit
even though the prediction is inside of the LMA-MSW region, 
giving a contribution to our goodness of fit of \mbox{\rm g.o.f.} 
$\approx 0.14$. Our CHOOZ angle also turns out to be 
about a factor of $2$ larger than the experimental limit at 
$90\%$ C.L., delivering another contribution of \mbox{\rm g.o.f.} 
$\approx 0.14$. In summary our predictions for the neutrino sector 
agree extremely well with the data, giving a contribution of only 
0.34 to \mbox{\rm g.o.f.} while the charged fermion sector contributes 
3.29 to \mbox{\rm g.o.f.}.

\subsection{$CP$ violation}
\indent\ Since we have taken our random couplings to be -- 
whenever allowed -- \underline{complex} we have order of 
unity or essentially maximal $CP$-violation so a unitary 
triangle with angles of order one is a success of our 
model. After our fitting of masses and of mixings 
we can simply predict order of magnitudewise of $CP$-violation in 
\mbox{\it e.g.} $K^0\! - \!\bar{K^0}$ decay or in CKM and MNS 
mixing matrices in general. 

The Jarlskog area $J_{\scriptscriptstyle CP}$ provides a measure of the 
amount of $CP$ violation in the quark sector~\cite{cecilia} and, 
in the approximation of setting cosines of mixing angles to unity, 
is just twice the area of the unitarity triangle:
\begin{equation}
  \label{eq:jarkskog}
  J_{\scriptscriptstyle CP}=V_{us}\,V_{cb}\,V_{ub}\,\sin \delta \,,
\end{equation}
where $\delta$ is the $CP$ violation phase in the CKM matrix.
In our model the quark mass matrix elements have random phases, 
so we expect $\delta$ (and also the three angles $\alpha$, 
$\beta$ and $\gamma$ of the unitarity triangle) to be of 
order unity and, taking an average value of 
$|\sin\delta| \approx 1/2$, the area of the 
unitarity triangle becomes
\begin{equation}
  \label{eq:jarkskog*0.5}
  J_{\scriptscriptstyle CP}\approx \frac{1}{2}\,V_{us}\,V_{cb}\,V_{ub}\,.
\end{equation}
Using the best fit values for the CKM elements from 
Table~\ref{convbestfit}, we predict 
$J_{\scriptscriptstyle CP} \approx 3.1\times10^{-6}$ to be compared with 
the experimental value $(2-3.5)\times10^{-5}$. 
Since our result for the Jarlskog area  
is the product of four quantities, we do not expect the 
usual $\pm64\%$ logarithmic uncertainty but rather 
$\pm\sqrt{4}\cdot64\%=128\%$ logarithmic
uncertainty. This means our result deviates from the 
experimental value by 
$\ln (\frac{2.7 \times 10^{-5}}{3.1 \times 10^{-6}})/1.28$ 
= 1.7 ``standard deviations''. 

The Jarlskog area has been calculated from the best fit parameters in 
Table 2, it is also possible to calculate them directly while 
making the fit. So we have calculated $J_{\scriptscriptstyle CP}$ 
for $N=10,000$ complex order unity random 
number combinations. Then we took the logarithmic average 
of these $10,000$ samples of $J_{\scriptscriptstyle CP}$
and obtained the following result:
\begin{eqnarray}
  \label{eq:jcpabsm}
   J_{\scriptscriptstyle CP}&=& 3.1\times 10^{-6} \,,
\end{eqnarray}  
in good agreement with the values given above.

\subsection{Neutrinoless double beta decay}
\indent\ Another prediction, which can also be made from this model, is 
the electron ``effective Majorana mass'' -- the parameter in  
neutrinoless beta decay -- defined by: 
\begin{equation}
\label{eq:mmajeff}
\abs{\sVEV{m}} \equiv \abs{\sum_{i=1}^{3} U_{e i}^2 \, m_i} \,,
\end{equation}
where $m_i$ are the masses of the neutrinos $\nu_i$ 
and $U_{e i}$ are the MNS mixing matrix elements for the 
electron flavour to the mass eigenstates $i$. We can 
substitute values for the neutrino masses $m_i$ from 
eqs.~(\ref{eq:neutrinomass1}-\ref{eq:neutrinomass3}) and for the 
fitted neutrino mixing angles from Table~\ref{convbestfit} into 
the left hand side of 
eq.~(\ref{eq:mmajeff}). 
As already mentioned, the $CP$ violating phases in the MNS mixing 
matrix are essentially random in our model. So we combine the 
three terms in eq.~(\ref{eq:mmajeff}) by taking the square root of 
the sum of the modulus squared of each term, which gives
our prediction:
\begin{equation}
  \label{eq:meffresult}
  \abs{\sVEV{m}} \approx 3.1\times 10^{-3}~~\mbox{\rm eV}\,.
\end{equation}

In the same way as being calculated the Jarlskog area we
can compute using $N=10,000$ complex order unity random 
number combinations to get the $\abs{\sVEV{m}}$. Then we 
took the logarithmic average 
of these $10,000$ samples of $\abs{\sVEV{m}}$ as 
usual: 
\begin{eqnarray}
    \abs{\sVEV{m}}&=& 4.4\times 10^{-3}~~\mbox{\rm eV}\,.
\end{eqnarray}  
This result does not agree with the central value of recent 
result -- ``evidence'' -- from the
Heidelberg-Moscow collaboration~\cite{evidence}.

\section{Baryogenesis via Lepton Number Violation}
\indent\ Having now a well fitted model giving orders of magnitude 
for all the Yukawa couplings and having the see-saw mechanism,
it is obvious that we ought to calculate the amount of baryons $Y_B$
relative to entropy being produced via the Fukugita-Yanagida 
mechanism~\cite{FY}. According to this mechanism the decay of the 
right-handed 
neutrinos by $CP$-violating couplings lead to an excess of the $B-L$
charge (meaning baryon number minus lepton number), the relative excess
in the decay from Majorana neutrino generation number $i$ 
being called $\epsilon_i$. This 
excess is then immediately -- and continuously back and forth -- being
converted partially to a baryon number excess, although it starts out 
as being a lepton number $L$ asymmetry, since the right-handed 
neutrinos decay to leptons and Weinberg-Salam Higgs particles.   
It is a complicated discussion to estimate to what extend the 
$B-L$ asymmetry
is washed out later in the cosmological development, but our estimates 
goes that there is not enough baryon number excess left to fit the 
Big Bang development at the stage of formation of the light elements 
primordially (nuclearsynthesis). 

Recently we have, however, developed a modified version~\cite{NTfur} 
of our model -- only deviating in the right-handed sector -- characterized 
by changing the quantum numbers assumed for the see-saw scale producing
Higgs field $\phi_{B-L}$ in such a way that the biggest matrix elements in 
the right-handed mass matrix (eq.~\ref{Mmajo}) becomes the pair of 
-- because of the symmetry -- identical off diagonal 
elements (row, column)=(2,3) and (3,2).
Thereby we obtain two almost degenerate right-handed neutrinos and that helps
for making the $B-L$ asymmetry in the decay bigger. In this modified model 
that turns out to fit the rest of our predictions approximately equally well
or even better we then get a very satisfactory baryon number relative to 
entropy prediction
\begin{equation}
Y_B \approx  2.5 \times 10^{-11}\,. 
\end{equation}   
 In the same time as making this modification of the $\phi_{B-L}$ 
quantum numbers we also made some improvements in the calculation 
by taking into account
the running of the Dirac neutrino Yukawa couplings from the Planck scale to 
the corresponding right-handed neutrino scales. Also, we calculated 
more accurate dilution factors than previous our work~\cite{NT2}. However,
foregoing work was based on the mass matrices which predicted
the SMA-MSW, so we must investigate the baryogenesis using the
present mass matrices, of course, with the modified right-handed
Majorana mass matrix.

\section{Conclusion}
\indent\ We have reviewed our model which is able to 
predict the experimentally favored LMA-MSW 
solution rather than the SMA-MSW solution for solar neutrino 
oscillations after careful choice of the $U(1)$ charges for the
Higgs fields causing transitions between 1st and 2nd 
generations. However, the fits of charged lepton quantities
become worse compare to our ``old'' model that can predict
SMA-MSW solar neutrino solution. On the other hand, we 
now can fit the neutrino quantities very well: the price 
paid for the greatly improved neutrino mass matrix fit -- the 
neutrino parameters now contribute only very little to the 
\mbox{\rm g.o.f.} -- is a slight deterioration in the fit to the 
charged fermion 
mass matrices. In particular the predicted values of the quark masses 
$m_d$ and $m_c$ and the Cabibbo angle $V_{us}$ are reduced compared to 
our previous fits. However the overall fit agrees with the seventeen 
measured quark-lepton mass and mixing angle parameters in Table 2 
within the theoretically expected uncertainty~\cite{FF} of about 
$64\%$; it is a perfect fit order of magnitudewise. 
It should be remarked that our model provides 
an order of magnitude fit{}/{}understanding of all the effective
Yukawa couplings of the Standard Model and the neutrino 
oscillation parameters in terms of only $6$ parameters -- the Higgs 
field vacuum expectation values!

\section*{Acknowledgments}
\indent\ We wish to thank the organisers of the
8th Adriatic Meeting for the wonderful 
organisation and for the hospitality extended to us
during the symposia. We thank 
to the Volkswagenstiftung for financial support. 
Y.T. wishes to thank S.~Koch for friendship and afternoon swimming 
lessons at Adriatic beach.


\end{document}